\documentclass[pra,aps, twocolumn, english]{revtex4}
\usepackage{graphicx}
\usepackage{amssymb, amsmath, color, rotating}

\begin{document}

\title{Domain wall dynamics in a two-component Bose-Mott insulator}

\author{Stefan S. Natu}
\email{ssn8@cornell.edu}
\author{Erich J. Mueller}
\email{em256@cornell.edu}
\affiliation{Laboratory of Atomic and Solid State Physics, Cornell University, Ithaca, New York 14853, USA.}

\begin{abstract}
We model the dynamics of two species of bosonic atoms trapped in an optical lattice within the Mott regime by mapping the system onto a spin model.  A field gradient breaks the cloud into two domains. We study how the domain wall evolves under adiabatic and diabatic changes of this gradient.  We determine the timescales for adiabaticity, and study how temperature evolves for slow ramps. We show that after large, sudden changes of the field gradient, the system does \textit{not} equilibrate on typical experimental timescales.  We find interesting spin dynamics even when the initial temperature is large compared to the super-exchange energy.   We discuss the implication of our results for experiments wishing to use such a two-component system for thermometry, or as part of a cooling scheme.
\end{abstract}
\maketitle

\section{Introduction} A major frontier in ultracold-atom physics is the search for novel low temperature ($T$) phases of matter. One well established success is the observation of Mott insulating shells of bosons in an optical lattice\cite{Greiner, chin2}.  More recently J\"ordens \textit{et al.} \cite{jordens} successfully observed the related Mott-metal crossover in a gas of fermionic $^{40}$K atoms in an optical lattice.  Much interest is now focussed on the magnetic properties of these insulators, which set in at the super-exchange scale $T \sim t^{2}/U$, where $t$ is the tunneling amplitude, and $U$ is the on-site interaction. For example, Trotzky \textit{et al.} \cite{trotzky} observed physics of super-exchange in a double well setup. This physics has not yet been seen in an extended system. The major impediments to these studies center around cooling, thermometry and equilibration. By considering an experimentally relevant setup involving two-component bosons, we are able to better characterize these impediments and explore ways to overcome them. We identify time-scales for adiabaticity, and the limits on the use of adiabatic demagnetization for cooling. 

The problem of cooling atoms in an optical lattice is particularly rich. It has been the focus of a large number of theoretical works: for example
M. Popp \textit{et al.},  B. Capogrosso-Sansone \textit{et al.},  J-S. Bernier \textit{et al.} and T. L. Ho \textit{et al.} \cite{Popp, Sansone, Bernier, jason2} considered separating the system into entropy rich and poor regions, and subsequently evaporating away atoms in the entropy-rich regions. The latter authors \cite{jason3} also proposed a scheme where one reduces the entropy per particle in the region of interest by introducing a deformation in the confining trapping potential as well as pointing out possible issues regarding heating while loading into an optical lattice \cite{jason1}. Bezett and Blakie \cite{Blakie} have performed a study of the adiabatic pathways for cooling fermions loaded into a three-dimensional optical lattice. J. Catani \textit{et al.} \cite{Catani} proposed using species-selective trapping potentials to transfer entropy from the target species to the auxillary species.  Additionally, F.Werner \textit{et al.} \cite{werner} have proposed using an analog of Pomeranchuk cooling in He-$3$ to adiabatically cool while loading a two-component Fermi gas in an optical lattice. 

Our study has relevance to all of these protocols. The key feature of each of these ideas is that they rely on \textit{adiabaticity}. Among the questions we address is: how slowly an external tunable parameter should be changed in order to maintain thermal equilibrium throughout? Not surprisingly, we find in the regime where super-exchange is important, the time-scale for adiabaticity scales as the inverse of the super-exchange energy.

In addition to the issues of cooling and adiabaticity, a large amount of work has been done on \textit{thermometry} \cite{Weld, jason4, demarco}. Recently D. Weld \textit{et al.} have suggested that a two-component Bose gas in the Mott regime can be used as an accurate thermometer. By displacing the potentials for the two species, one creates a domain wall. The width of that domain wall ($w$) is proportional to the temperature. It is this two-component Bose system  that we study here. The question of thermometry is intimately connected to the question of adiabaticity and equilibration. The system needs to be in thermal equilibrium for temperature to be well defined. 

Finally, our studies of the dynamics of this system fit in with ongoing works on ultra-cold atoms far from equilibrium \cite{cornell, zwierlein, thomas, sadler, kinoshita, rigol, demler3, kollath, muramatsu, demler2, cardy, sachdev, polkovnikov}.  Cold atoms systems are readily isolated from their environment, and their parameters may be rapidly tuned.  This is largely unexplored territory, and organizing principles have only been found for a small subset of the possible phenomena.  The response of this bosonic system to a changing field gradient is one of the more controlled avenues for exploring non-equilibrium physics.

This paper is organized as follows: In Sec.II we discuss the experiment of Weld \textit{et al.} and introduce our model. In Sec. III we discuss the thermodynamics of the model and describe our numerical procedure for calculating dynamics. In Sec. IV we determine the time-scales required for maintaining global equilibrium, and in Sec. V we discuss the idea of adiabatic demagnetization applied to our existing setup.
In Sec. VI  we show that effects of superexchange can be manifested even at larger temperatures. 

\section{Spin-gradient thermometry}
In \cite{Weld}, Weld, Medley, Miyake, Hucul, Pritchard and Ketterle created a two-component Mott insulator out of two hyperfine states of $^{87}$Rb in a cubic lattice. They worked well within the Mott regime using a lattice with a depth $V_{0} = 14.5 E_{R}$ ($t/U = 0.024$). At unity filling factor, this should be contrasted with lattice depth of $13.1 E_{R}$ ($t/U = 0.034$) for the superfluid to Mott-insulator transition in three dimensions \cite{prokofiev}.
Here $E_{R} = \frac{2\pi^{2}}{m\lambda^{2}}$ is the lattice recoil energy where the lattice spacing is $\lambda$, the atomic mass is $m$ and $\hbar$ has been set to $1$. 

The atoms are trapped in two internal states, treated here in a pseudo-spin-$\frac{1}{2}$ framework as $\sigma = \{\uparrow = |F=1, m_{F}=-1\rangle, \downarrow = |2, -2\rangle\}$. In terms of the creation ($a^{\dagger}_{\sigma, i}$) and annihilation ($a_{\sigma, i}$) operators for the respective spins at site $i$, the lattice Hamiltonian takes the two-component Bose-Hubbard form: 
\begin{eqnarray} \label{eq:1}
&&\mathcal{H}_{int}  = -\sum_{\langle ij\rangle,\sigma}t_{\sigma}(a^{\dagger}_{i\sigma}a_{j\sigma} + h.c) +\\\nonumber
&&\frac{1}{2}\sum_{\sigma, i}U_{\sigma\sigma}n_{i\sigma}(n_{i\sigma} -1) + \sum_{i\sigma}U_{\uparrow\downarrow}n_{i\uparrow}n_{i\downarrow} + \sum_{i}V_{i\sigma}a^{\dagger}_{i\sigma}a_{i\sigma} \\\nonumber
\end{eqnarray}
where $\langle ij \rangle$ refers to nearest-neighbor sites separated by the lattice spacing $a$, $U_{\sigma \tau}$ denotes the on-site repulsion between atoms in state $\sigma$ and $\tau$, and $t_{\sigma}$ is the tunneling rate. The expressions for the spin-dependent tunneling energies can be found in Ref.\cite{demler}. For the particular case of $^{87}$Rb considered here, $t_{\uparrow} \approx t_{\downarrow}$, $U_{\uparrow\uparrow} \approx U_{\downarrow\downarrow} \approx U_{\uparrow\downarrow} = U$. 

The external potential $V_{\sigma}$ is comprised of two terms. Red-detuned laser beams provide a Gaussian trap which is almost indistinguishable for the spin-states in consideration. In addition, there is a spin-dependent term, arising from a dynamically tunable external magnetic field, which varies linearly in space, resulting in a time-dependent Stern-Gerlach separation of the two spin states. 

The lowest excitations about the Mott insulator ($U \gg t$) do not arise from single particle tunnelings ($\sim t_{\sigma}$), but consist of exchanging $\uparrow$ and $\downarrow$ particles between neighboring sites. Second order perturbation theory shows that the energy cost for such a process is $\sim t^{2}/U$ \cite{demler, kuklov}. 

In the experiment the center of the cloud is a Mott insulator with one particle per site. One describes this region via an effective Hamiltonian: 
\begin{equation}\label{eq:2}
\mathcal{H}_{eff} = -J \sum_{\langle ij\rangle}\hat{\textbf{S}}_{i}\cdotp\hat{\textbf{S}}_{j} - \Delta\mu|B'|a\sum_{i_{Z}}i\hat{S}_{z},
\end{equation}
where  we use spin-$\frac{1}{2}$ operators at every site ($i$): $\vec{S}_{i} = \frac{1}{2} \sum_{\sigma\sigma^{'}} a^{\dagger}_{\alpha\sigma}\vec{\sigma}_{\sigma\sigma^{'}}a_{\alpha\sigma}$ where $\vec{S}_{i} = \{S_{xi},S_{yi},S_{zi}\}$. The $Z$ coordinate of lattice site $i$ is $I_{Z}$. Throughout, we use lower case alphabets ($x, y, z$) to denote spin-space and upper case alphabets ($X, Y, Z$) to denote real-space. The super-exchange energy is $J = 2t_{\uparrow}t_{\downarrow}/U$ and $\Delta\mu = \mu_{\uparrow} - \mu_{\downarrow} = \mu_{B}$, is the difference between the magnetic moments of the two states. The lattice spacing is $a$ and the magnetic field gradient is $B^{'}$. If a different atom was used, one may have to correct these expressions for the species dependence of the tunneling, and on-site energies. For a lattice depth of $14.5 E_{R}$, the super-exchange scale is $J \sim 50$pK.  There are two dimensionless numbers in the problem: $\beta J$ and $\beta E_{mag}$ where $E_{mag} = \Delta\mu_{B}|B^{'}|a$ and $\beta = 1/k_{B}T$. for time-independent magnetic fields, the Hamiltonian conserves $|\textbf{S}^{2}|$ at each site, total $S_{z}$ and the total energy. 

The standard Weiss mean-field theory yields  \cite{chaikin} 
\begin{equation}\label{eq:3} 
\langle S_{z i} \rangle = \tanh(\beta h_{i})
\end{equation} 
where $h_{i} = \frac{1}{2}(J\sum_{j \in nn~i}\langle S_{z j}\rangle - \Delta\mu B(i))$, with the sum over all nearest neighbor spins. Weld \textit{et al.} \cite{Weld} extract a temperature by fitting their experimental profiles to Eq.~\ref{eq:3}, using the independent site approximation, $J=0$. This technique appears to be one of the most powerful thermometry strategies applicable to these temperatures. The lowest measured temperatures in this experiment are $T \sim 1$ nK. The super-exchange energy scale is still a factor of $20$ smaller than this, justifying the single site approximation. 
 
For this technique to provide accurate thermometry, one must choose the field gradient appropriately. If the field gradient is too large, one cannot resolve the domain wall. If it is too small, the finite cloud size becomes important.

\begin{figure}[tbp]
\begin{picture}(230, 75)(10, 10)
\put(5, 10){\includegraphics[scale=0.35]{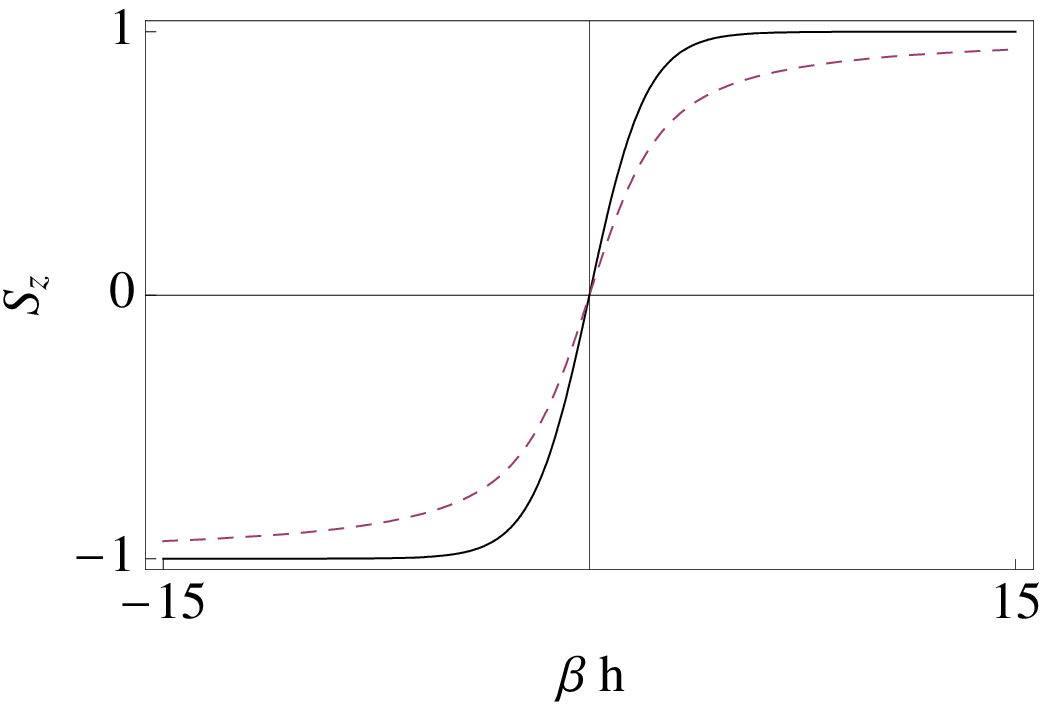}}
\put(120, 10){\includegraphics[scale=0.4]{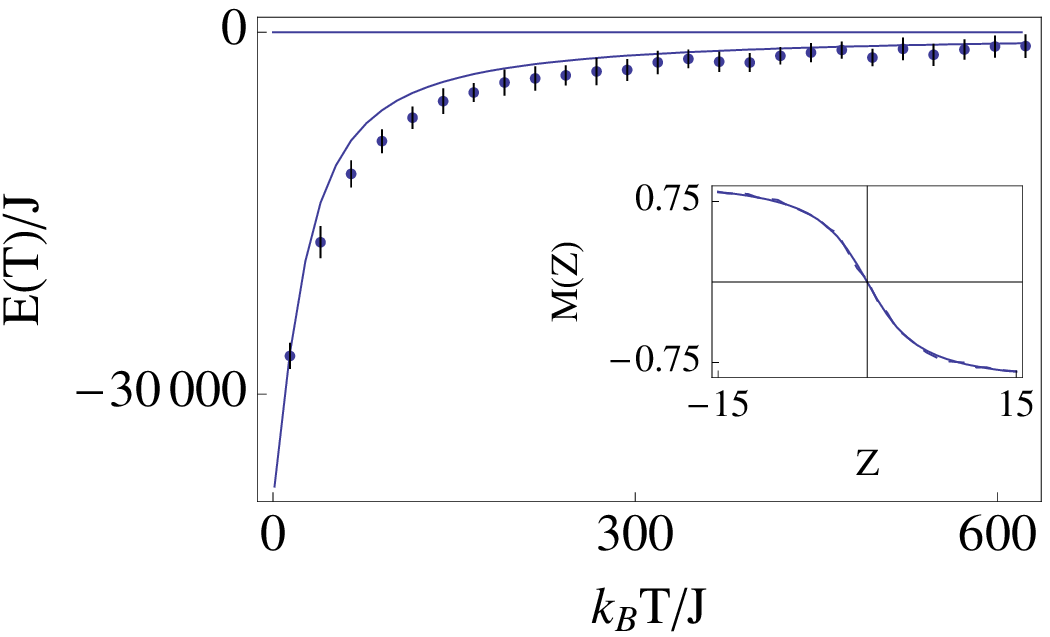}}
\end{picture}
\caption{\label{fig:-1} (Color Online) \textbf{Comparison between mean-field models and Monte-Carlo:} (Left): magnetization ($S_{z}$) gradient as a function of $\beta h$ from Ising spin-$\frac{1}{2}$ (solid), and classical (dotted) mean-field theories. At $|B'| = 2$G/cm, and $T = 1$nK, $(E_{mag} = \mu_{B}|B^{'}|a)/k_{B}T \sim 15$. (Right): A comparison of the energy as a function of temperature at $E_{mag} \sim J$ ($|B^{'}| = 0.01$G/cm), using classical mean-field theory (solid) and Metropolis Monte-Carlo (data points with error bars). Inset shows typical magnetization profiles obtained from these theories for the same gradient and $T =1$nK. Position $Z$ is measured in units of the lattice spacing $a$. The Monte-Carlo and mean-field curves are indistinguishable.}
\end{figure}

At the temperatures of interest, there are two contributions to the entropy density: (i) spin entropy  $S_{spin} \le \log(2)$ per site and (ii)  entropy from particle-hole excitations. At the final lattice depth in \cite{Weld} of $14.5 E_{R}$, we have a $t/U \sim 0.024$, and a Mott gap $\Delta \sim 15.5 t$. At $T = 1$nK $\sim t$, particle-hole excitations are suppressed by $\exp(-\Delta/T)$, and at the center of the trap, nearly all of the entropy of the system is contained in the spin-excitations. Given that the Mott gap goes to zero at the edge of the cloud, a separate entropy reservoir exists here. The thermal conductivity of the Mott state is so low that these reservoirs can be considered decoupled.

This poor thermal contact between the spin and particle degrees of freedom means that the domain wall size is a probe of the local spin temperature and may give very little information of the temperature at the edge of the cloud. Depending on the physics one is interested in, this may be beneficial or detrimental. Similarly, if this ``thermometer" is put in contact with another system of interest, it may be difficult for the two systems to thermalize with one another.  Essentially one needs to engineer efficient mechanisms for spin exchange.  Longer range spin interactions (such as dipolar interactions) would help in this regard.

\section{Dynamics and Thermodynamics}
Starting with the Heisenberg Hamiltonian Eq.~(\ref{eq:2}), we use the spin-commutation relations to obtain the equations of motion for the spin operator $\textbf{S}_{i} = \{S_{x},S_{y},S_{z}\}$. We then make a classical spin approximation, replacing these operators with their expectation value, and sequentially evolve the sites using a split-step approach. We verify that our timesteps are small enough that the total energy (for time-independent $B$), total $S_{z}$, and $|\textbf{S}|^{2}$ are conserved to within $0.1$ percent. 

In order to compare with our dynamics simulations, which treat the spins within a classical framework, we replace Eq. \ref{eq:3} with the the analogous mean-field theory for the classical model \cite{chaikin}:
\begin{equation}\label{eq:4}
M = \langle S_{z i} \rangle = \langle \cos(\theta) \rangle = \coth{\beta h_{i}} - \frac{1}{\beta h_{i}}.
\end{equation} 
where $h_{i}  = -2J\sum_{\langle j \rangle}S_{z j} - \Delta\mu B(i)$ is the effective magnetic field at site $i$. 

\begin{figure}[tbp]
\begin{picture}(200, 70)
\put(-20, 0){\includegraphics[scale=0.35]{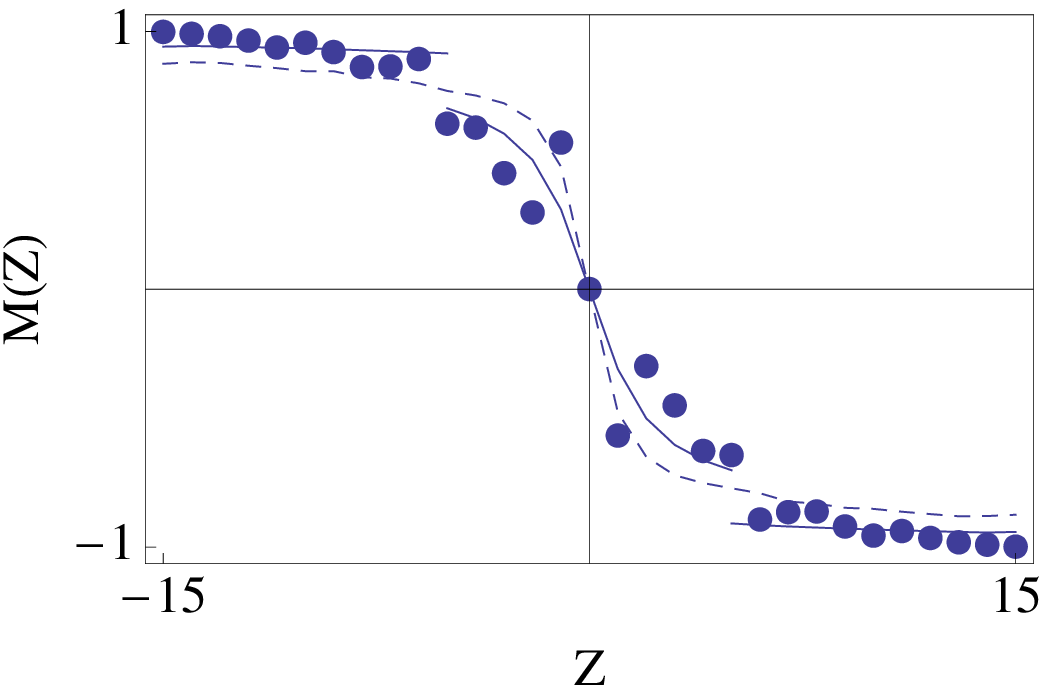}}
\put(110, 0){\includegraphics[scale=0.35]{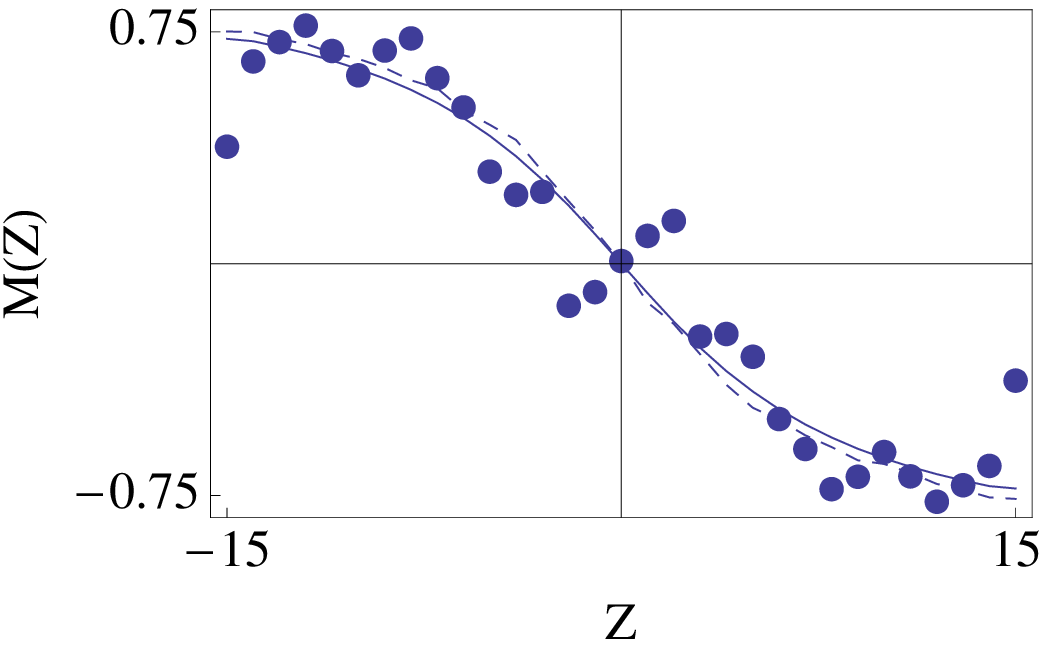}}
\end{picture}
\caption{\label{fig:-2} (Color Online) \textbf{Identifying time-scales for adiabaticity:} (Left) Time evolved magnetization profile (dots) after $\tau_{f} = 1$s starting with an initial field gradient of $0.25$G/cm to a final value of $0.01$G/cm at an initial temperature $T = 3$nK with rate $\Gamma = 20 J$.  The dots deviate considerably from the equilibrium profile from Monte-Carlo (dashed). The solid lines are obtained by fitting the center and wings to two different temperatures using Eq.\ref{eq:4}, $0.35$nK and $0.1$nK respectively. (Right) Same but with $\Gamma/J = 1$. This profile agrees well with Monte-Carlo and mean-field predictions for the equilibrium state at $0.35$nK. Position $Z$ is measured in units of the lattice spacing $a$.} \end{figure}

In Fig.\ref{fig:-1}(left) we compare the two mean-field models. At both small and large $\beta h$, Eqs. \ref{eq:3} and \ref{eq:4} agree.  The differences at intermediate $\beta h$ arise from the larger entropy of the classical model. Even though the experiments are better described by a quantum spin-$\frac{1}{2}$ model, we work with the classical model;  we do not have an efficient algorithm for the quantum model. At zero temperature, our approach can be viewed as a time-dependent variational ansatz, where we introduce $\psi = \Pi_{j}(\cos(\theta_{j}/2)\exp(i\phi_{j}/2)a^{\dagger}_{\uparrow j} + \sin(\theta_{j}/2)\exp(-i\phi_{j}/2)a^{\dagger}_{\downarrow j}$, where $\theta_{j}$ and $\phi_{j}$ represent the direction of the classical spin on site $j$. This product ansatz should capture most of the gross dynamics, while drastically reducing the numerical complications.

To capture thermodynamics beyond mean-field theory, we use a classical Metropolis Monte-Carlo algorithm. In Fig.~\ref{fig:-1}(right) we compare the classical mean-field theory to the Monte Carlo results. Later we compare the profiles obtained from the non-equilibrium dynamics to those obtained from both the classical mean-field theory and the Monte-Carlo.

\section{Adiabaticity}

\begin{figure*}[tbp]
\begin{picture}(100, 110)
\put(75, -5){\includegraphics[scale=0.55]{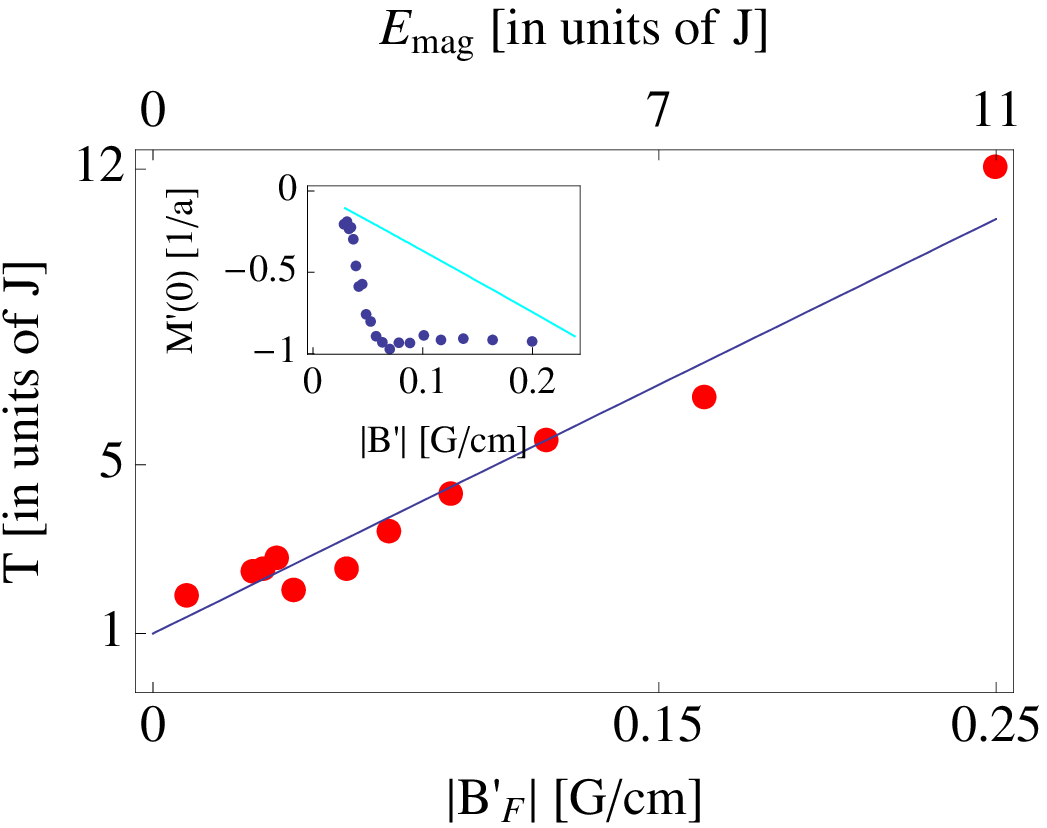}}
\put(-155, -5){\includegraphics[scale=0.55]{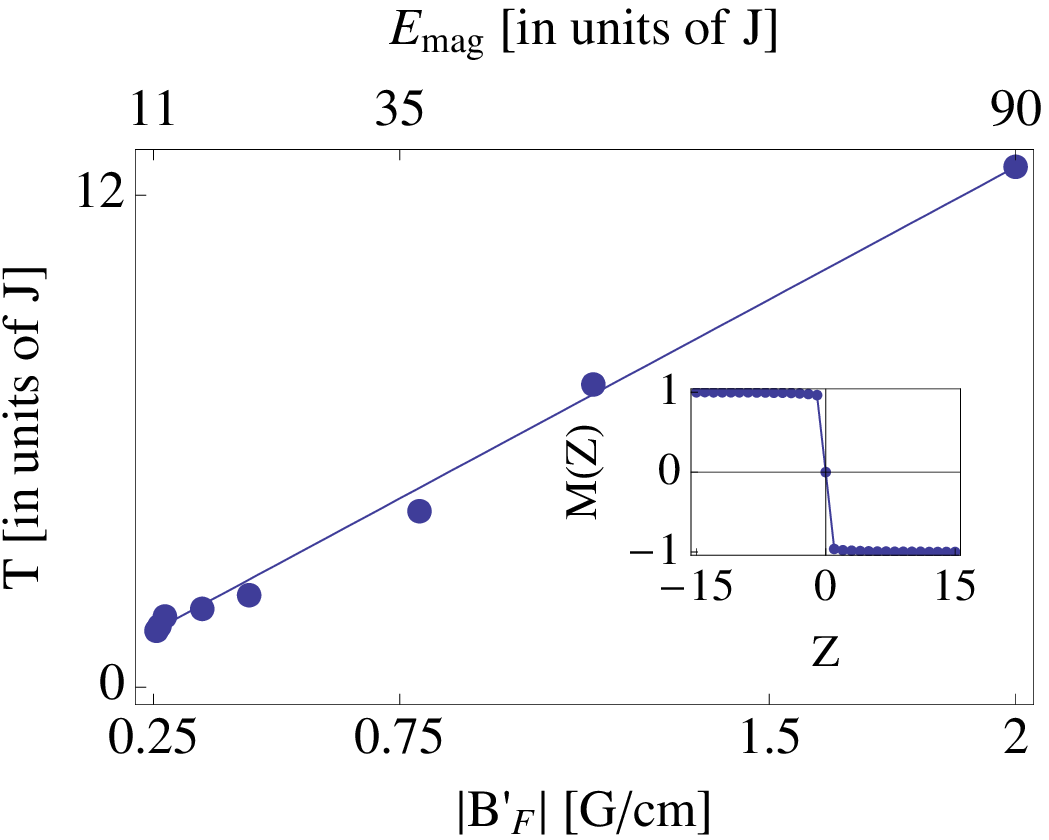}}
\end{picture}
\caption{\label{fig:-3} (Color Online) \textbf{Cooling via adiabatic demagnetization:} Left: Temperature evolution as a function of magnetic field gradient ($|B_{F}^{'}|(\tau)$) and $E_{mag} = \mu_{B}B^{'}a$ at various times $\tau$, from $|B_{initial}^{'}| = 2$G/cm to $|B_{F}^{'}(\tau = 1)| = 0.25$G/cm with $\Gamma/J = 20$. The temperature is extracted by fitting the spin density to the functional form in Eq.~(\ref{eq:4}). Inset shows a comparison between the magnetization profiles from the time-evolution and Eq.~\ref{eq:4} at $T=0.1$nK. Position $Z$ is measured in units of the lattice spacing $a$. Right: Same but with $|B_{initial}^{'}| = 0.25$G/cm to $|B_{final}^{'}| = 0.01$G/cm with $\Gamma/J = 1$, using Monte-Carlo. Lines through the data are guides to the eye. Inset shows the  gradient of the central magnetization ($\propto 1/w$) as a function of the magnetic field gradient from the time-evolution (points) and Eq.~\ref{eq:4} (solid line) in the independent site approximation $J=0$.}
\end{figure*}

Here we consider the question of exactly how slowly one must change parameters in order to remain in thermal equilibrium. Understanding this timescale is essential for various proposed cooling schemes. We will focus on time-dependent field gradients. There are two distinct regimes to consider:  (i) $E_{mag} \gg J$ - super-exchange is largely irrelevant, and one requires the field gradient to be lowered at a rate small compared to $\omega_{L}$ the local Larmor precession frequency and (ii) $E_{mag} \le J$ - the gradient should be lowered at a rate small compared to $J$.

We pick a functional form for the magnetic field gradient $B^{'}|_{\tau} = \frac{B^{'}_{0} - B^{'}_{\tau_{f}}}{(1 + \tau \Gamma)^{2}} + B^{'}_{\tau_{f}}$, where the rate $\Gamma$, proportional to $\dot B^{'}/B^{'}$, characterizes the adiabaticity of the evolution: $\Gamma/J \rightarrow \infty$ indicates a sudden quench, while for $\Gamma/J \sim 1$, we expect the magnetization profile to adiabatically follow the equilibrium profile. 

To illustrate this we lower the magnetic field gradient from an initial value of $0.25$G/cm, ($E_{mag} \sim 25 J$) to $B^{'}|_{\tau_{f}} = 0.01$G/cm ($E_{mag} \sim J$) at various rates. The initial state is chosen to obey Eq.~\ref{eq:4} with a temperature of $T=3$nK and with uncorrelated random transverse spins. I.E. we choose $\{S_{x}, S_{y} \} = \sqrt{1- (S_{z})^{2}}\{\cos(\phi), \sin(\phi)\}$, where the azimuthal orientation $\phi$ is allowed to be random at each site. In order to more accurately reflect experimental conditions, we introduce spatially random noise in the ambient magnetic field gradient at the $10$ percent level. We ensemble average over $10$ different noise profiles.  We consider a rectangular geometry with $5 \times 5 \times 30$ lattice sites in the $X-Y$ and $Z-$ directions respectively.

In Fig.~\ref{fig:-2}, we plot the time-evolved magnetization profiles (dots) in the direction of the field gradient, after ensemble averaging and averaging over the transverse directions, at $\tau_{f}$ ($\sim 1s$) for $\Gamma/J \sim 20$ (left) and $\Gamma/J \sim 1$ (right). By comparing the energy of the time-evolved profile to the classical Monte-Carlo simulation, we get an indication of how well thermalized the system is. For example, after the more rapid change in the field gradient, we find that the center of the cloud is fit better to a higher temperature than the wings. On the other hand, after the slow ramp, the coarse-grained magnetization profile and the final temperature agree well with the equilibrium expectations.

Both of these ramps were slow compared to the Larmor precession frequency, so local equilibrium was maintained.  However the timescale for global equilibration is set by super-exchange. The behavior seen in Fig.\ref{fig:-2} (left) is reminiscent of that observed by Hung \textit{et al.} \cite{gemelke} in a single component Bose system, upon turning on the lattice. They too observed that different parts of the cloud were fit better by different temperatures. In order to stay close to equilibrium, it is crucial that external parameters be changed slowly compared to the timescale for rearranging the state.

\section{Adiabatic cooling}

Having established the timescale for adiabaticity, we now critically evaluate how adiabatically changing the field gradient can lead to cooling. This is analogous to the adiabatic demagnetization technique used in solid state physics. 

Adiabatic cooling is based upon the following idea: For a large initial field gradient, the spin entropy is largely contained in the domain wall, while in the bulk the spins are all either pointing up or down. As we lower the field gradient \textit{adiabatically}, the entropy of the system stays constant, and remains concentrated at the domain wall.  This effectively means that the size of the domain wall does not change: the entropy contained in the domain wall is $\log(2)$ times the number of spins in it. Since the domain wall width is proportional to the ratio of the temperature to the magnetic energy $w \propto T/(\Delta\mu \frac{\partial B}{\partial x})$, one concludes  that if $w$ stays fixed, $T$ must go down. Hence lowering the field gradient cools the system. Note that when $E_{mag} \sim J$, this argument breaks down as the domain wall profile is set largely by the super-exchange energy, rather than $E_{mag}$.

In Fig. \ref{fig:-3} (left) we adiabatically lower the field gradient from $2$G/cm to $0.25$G/cm, at a rate of $\Gamma/J = 20$, starting with an initial temperature of $T =1$nK. As we evolve the system, we extract the temperature by comparing the profiles to our Monte Carlo simulations. Due to slight deviations from adiabaticity, this eight fold decrease in the field gradient only results in a six-fold decrease in the temperature. The inset shows the magnetization profile at the final time along with the best fit Monte-Carlo profile. 

Fig, \ref{fig:-3} (right) illustrates what happens when $E_{mag}$ is smaller: cooling stops when $E_{mag} \sim J$. The inset shows how the slope of the domain wall varies with field gradient. For this simulation, we worked with a much slower ramp taking $\Gamma = J$. We found that faster ramps were not adiabatic.

\section{Super-exchange effects at $T > J$}

\begin{figure}[tbp]
\begin{picture}(300, 80)
\put(140, 5){\includegraphics[scale=0.35]{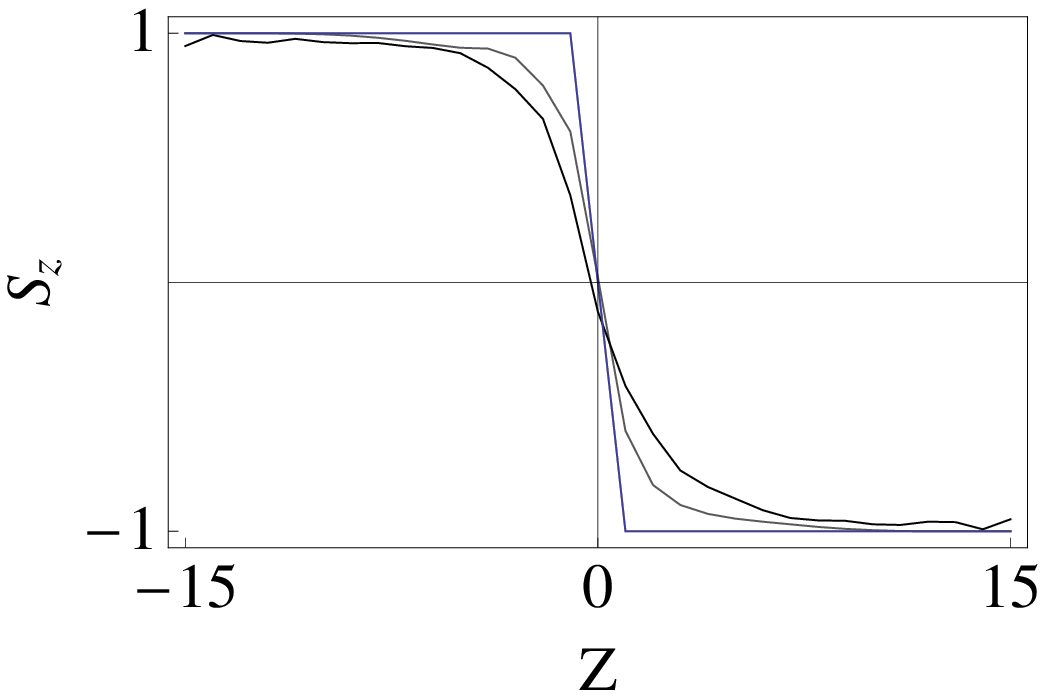}}
\put(0, -5){\includegraphics[scale=0.45]{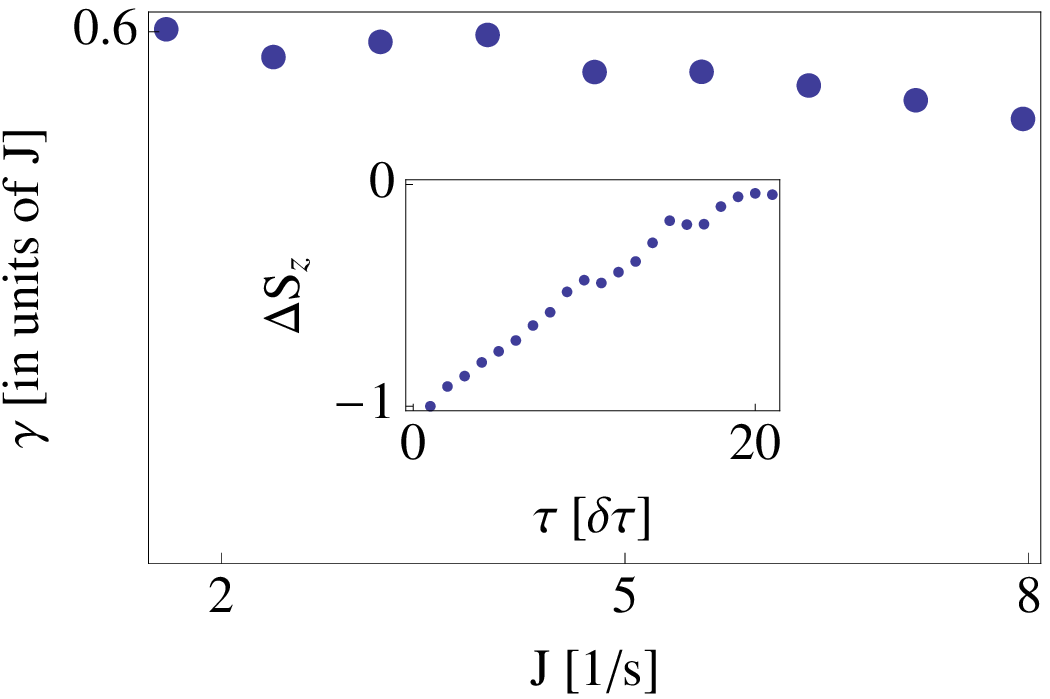}}
\end{picture}
\caption{\label{fig:-4} (Color Online) \textbf{Sudden Quench:} Left: Rate ($\gamma_{z} = \partial\Delta S_{z}/\partial\tau$) for different values of $J$, upon quenching from an initial state at $T  =1$nK, taking $B_{i}^{'}(\tau) = 2\Theta(-\tau)$G/cm. Here  $\Delta S_{z}$ is the difference between the magnetization at the center and one lattice site away. Inset shows how $\Delta S_{z}$ changes in time, where we use $\delta \tau J = 0.025$. Right: Typical magnetization profiles following the quench at $\tau = 0, 2J^{-1}, 4J^{-1}$, where flatter profiles indicate later times. Position $Z$ is measured in units of the lattice spacing $a$.} 
\end{figure}

As shown by Quantum Monte Carlo simulations, the spin-$\frac{1}{2}$ Heisenberg model has a ferromagnetic transition, in zero field at a temperature $T_{c} \sim 0.9J$ \cite{wessel}. Our classical model has a similar transition at $T_{c} = 1.5J$. At temperatures above $T_{c}$, the domain wall becomes arbitrarily large as $B^{'} \rightarrow 0$. Below $T_{c}$ the domain wall size will saturate.  In addition to these gross features, there may be interesting physics involving the transverse spins in the domain wall.  At high temperatures these are certainly disordered, while at low temperature it would be reasonable to expect them to be ordered (at least algebraically).

The excitations about the ferromagnet are low energy, quadratically dispersing, spin waves. At finite temperature these can be thought of as non-interacting particles (magnons) with an average density given by the Bose occupation factor. This picture breaks down near the Curie transition, as magnons can scatter off one another, and do not behave like free-particles. As shown by Hohenberg and Halperin \cite{halperin}, spin-waves in the paramagnetic phase are over-damped, and develop a sharp peak at $k=0$ at $T = T_{c}$.

\subsection{Domain wall dynamics}
Although the domain wall profile in Weld \textit{et al.} \cite{Weld} does not depend on $J$, it is superexchange that drives the spin dynamics.  Thus dynamics provides the clearest signature of magnetic interactions in the paramagnetic phase.  Here we explore the dynamics of the domain wall profile when the magnetic field gradient is suddenly turned off.  
Once the gradient is removed, the domain wall spreads out and flattens -- presumably moving towards the equilibrium state which is a homogenous paramagnet. The dynamics are driven by spin currents and the characteristic wave-length of the relevant magnons is at the scale set by the domain wall size.

We characterize the domain wall shape by the rate $\gamma_{z} = \partial\Delta S_{z}/\partial\tau$, where $\Delta S_{z}$ is the difference between the magnetization at the center, and one lattice site away. In Fig.~\ref{fig:-4} (left) we plot this quantity as a function of $J$, starting at $T  =1$nK.  We take $B_{i}^{'}(\tau) = 2 \Theta(-\tau)$G/cm  to be discontinuous: $\Theta(-\tau)$ is the Heaviside step function.  We see that $\gamma \propto J$, this is consistent with the fact that $J$ provides the scale for spin transport. In particular, spin wave theory would give this result. On the right we show the spreading of the domain wall in time ($\tau = 0, 2J^{-1}, 4J^{-1}$), where the later profiles are flatter. 

\begin{figure}[tbp]
\begin{picture}(200, 80)
\put(100, -5){\includegraphics[scale=0.4]{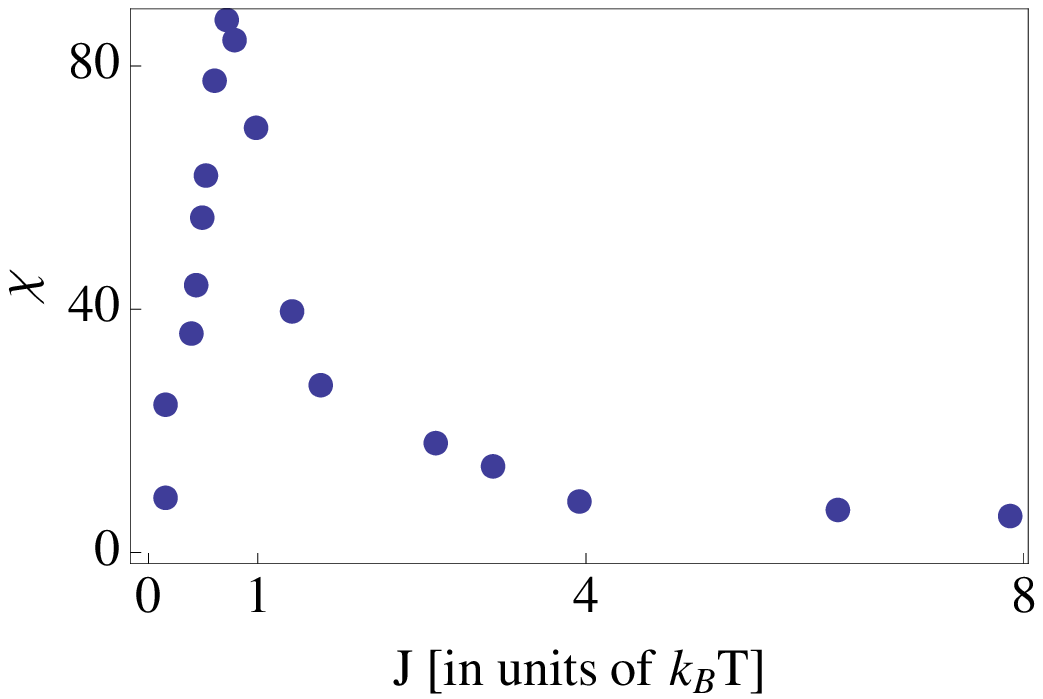}}
\put(-20, -5){\includegraphics[scale=0.4]{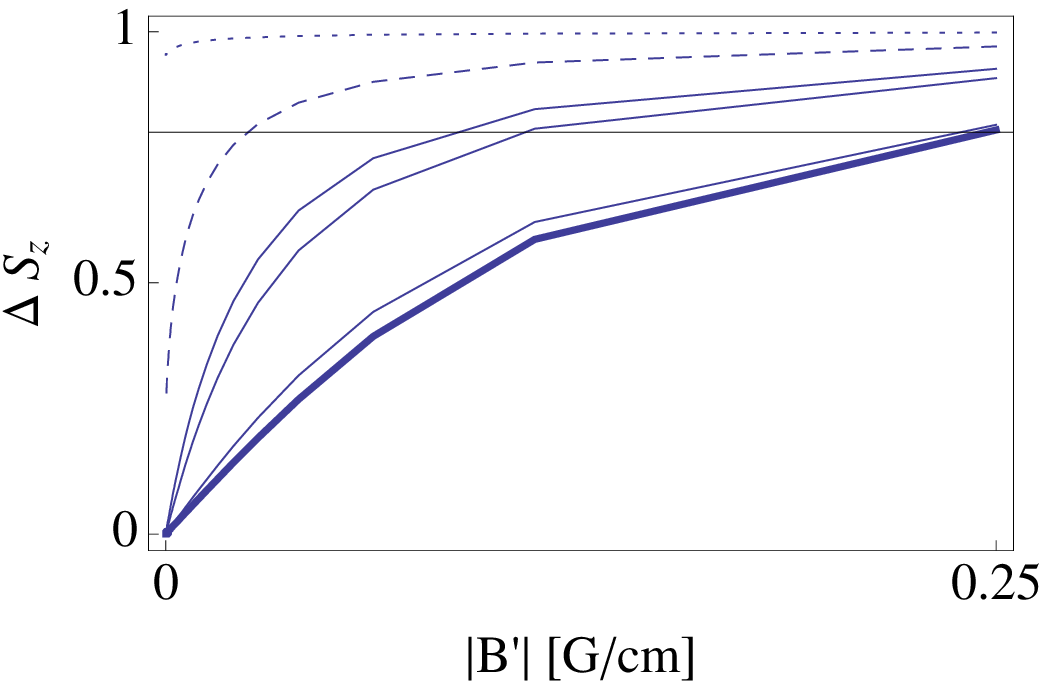}}
\end{picture}
\caption{\label{fig:-5} (Color Online) \textbf{Susceptibility:} Left: $\Delta S_{z}$ as a function of the field gradient at different temperatures using classical mean-field theory: $T > T_{c}$ (solid),  $T_{c}$ (dashed), and  $T \approx 0$ (dotted). Right: Susceptibility $\chi= \partial \Delta S_{z}/\partial B^{'}|_{B^{'}=0}$ in units of $1/$(G/cm) as a function of $\beta J$ from classical mean-field theory diverges with the mean-field exponent $\sim 1.5$ as $T \rightarrow T_{c}$}.
\end{figure}

The magnitude of $\gamma$ depends on the width ($w$) of the initial domain wall, which was roughly one lattice site in the above simulation. We find that doubling $w$, reduces $\gamma$ by $2$. This scaling arises because of the quadratic spin-wave dispersion. The group velocity $v(k) = d\epsilon/dk$ falls by a factor of $2$ when the wave-vector is halved. We find that for very large initial domain walls the dynamics becomes extremely slow. 

These dynamics give some of the same information that in a traditional condensed matter setting one would seek from transport measurements. Such transport measurements are extremely hard in cold atoms as there is no natural way to introduce leads. Here one instead needs to infer the spin current by observing the spin dynamics. 

\subsection{Signatures of the phase transition}

In Fig.~\ref{fig:-5} (left) we plot $\Delta S_{z}$ as a function of the field gradient at different temperatures using the classical mean-field model. The Curie transition is manifested by a divergence in the derivative of $\Delta S_{z}$ with respect to the field gradient at zero field gradient ($\chi= \partial \Delta S_{z}/\partial B^{'}|_{B^{'}=0}$). On the right we plot $\chi$ as a function of temperature, finding that as $T \rightarrow T_{c}$, it diverges with a mean-field exponent of $1.5$.

\section{Summary}
We have modeled the dynamics of the two-component Bose-Mott insulator by mapping it onto a spin model. By comparing semi-classical dynamics to classical Monte Carlo, and mean-field theory we extract the temperature of the system. We show that for experiments conducted in the Mott regime, the super-exchange sets the equilibration rate, and adiabaticity timescales. We find that adiabatic demagnetization in this system can lead to significant cooling and expect this cooling to be observed experimentally. Finally the slope of the domain wall as a function of the field gradient is an indicator of the Curie transition, as it exhibits a divergence at this point. We propose that observing domain wall dynamics is the most promising avenue for studying super-exchange physics in the paramagnetic phase. 

\section{Acknowledgements} SN would like to thank Stefan K. Baur, Kaden R. A. Hazzard and Mukund Vengalattore for insightful comments.  This work was supported by a grant from the Army Research office with funding from the DARPA OLE Program.


\begin{thebibliography}{99}

\bibitem{Greiner} M. Greiner, O. Mandel, T. Esslinger, T. W. H\"ansch and I. Bloch, Nature (London), \textbf{415} 39 (2002).

\bibitem{chin2} N. Gemelke, X. Zhang, C-L. Hung, C. Chin, Nature, \textbf{460}, 995 (2009).

\bibitem{jordens} R. J\"ordens, N. Strohmaier, K. G\"unter, H. Moritz, T.Esslinger, Nature, \textit{455} 204 (2008).

\bibitem{trotzky} S. Trotzky, P Cheinet, F. F\"olling, M. Feld, U. Schnorrberger, A. M. Rey, A. Polkovnikov, E. A. Demler, M. D. Lukin, I. Bloch, Science \textbf{319} 295 (2008). 

\bibitem{Popp} M.Popp, J.J.Garcia-Ripoll, K.G.Vollbrecht, J.I.Cirac, Phys. Rev. A \textbf{74} 013622 (2006).

\bibitem{Bernier} J.S.Bernier, C.Kollath, A.Georges, L. DeLeo, F. Gerbier, C. Salomon, M. K\"{o}hl, Phys. Rev. A \textbf{79}, 061601(R) (2009).

\bibitem{jason2} T.-L Ho, Q. Zhou, Proc. Natl. Acad. Sci. U.S.A \textbf{106} 6916 (2009).

\bibitem{Sansone} B.Capogrosso-Sansone, S. G. S\"oyler, N. Prokof'ev, and B. Svistunov, Phys. Rev. A \textbf{77} 015602 (2008).

\bibitem{jason3} T.-L. Ho, Q. Zhou, eprint arxiv:0911.5506. 

\bibitem{jason1} T. -L. Ho, Q. Zhou, Phys. Rev. Lett. \textbf{99} 120404 (2007).

\bibitem{Blakie} P.B.Blakie, A. Bezett, Phys Rev. A \textbf{71} 033616 (2005).

\bibitem{Catani} J.Catani, G. Barontini, G. Lamporesi, F.Rabatti, G.Thalhammer, F. Minardi, S. Stringari, and M.Inguscio, Phys. Rev. Lett. \textbf{103} 140401 (2009).

\bibitem{werner} F. Werner, O. Parcollet, A. Georges, and S. R. Hassan, Phys. Rev. Lett. \textbf{95} 056401 (2005).

\bibitem{Weld} D.M. Weld, P.Medley, H. Miyake, D. Hucul, D. E. Pritchard, and W. Ketterle, Phys. Rev. Lett. \textbf{103}, 245301 (2009).

\bibitem{jason4} Q. Zhou and T.-L. Ho, eprint arxiv:0908.3015

\bibitem{demarco} D. McKay, M. White and B. DeMarco, Phys. Rev. A \textbf{79} 063605 (2009).

\bibitem{prokofiev} B Capogrosso-Sansone, N.V.Prokof'ev, B.V.Svistunov, Phys. Rev. B. \textbf{75} 134302 (2007).

\bibitem{cornell} H.J.Lewandowski, D.M.Harber, D.L. Whitaker and E.A. Cornell, Phys. Rev. Lett., \textbf{88} 070403 (2002).

\bibitem{zwierlein} A. Sommer, A. Schirotzek, M. Ku and M. Zwierlein, (unpublished).

\bibitem{thomas} X.Du, L.Luo, B.Clancy and J.E. Thomas, Phys. Rev. Lett., \textbf{101} 150401 (2008). 

\bibitem{sadler} L.E.Sadler, J.M.Higbie, S.R.Leslie, M.Vengalattore and D.M.Stamper-Kurn, Nature Letters, \textbf{443} (2006).

\bibitem{kinoshita} T. Kinoshita, T.Wenger, D. Weiss, Nature (London), \textbf{440}, 900 (2006). 

\bibitem{rigol} M.Rigol, V. Dunjko and M. Olshanii, Nature, \textbf{452}, 854-858 (2008).

\bibitem{demler3} V.Gritsev, P.Barmettler, and E.Demler, eprint arXiv: 0912.2744.

\bibitem{kollath} C. Kollath, A. M. L\"auchli, and E. Altman Phys. Rev .Lett. \textbf{98}, 180601, (2007).

\bibitem{muramatsu} S. R. Manmana, S. Wessel, R. M. Noack, A. Muramatsu, Phys. Rev. Lett. \textbf{98} 210405 (2007).

\bibitem{demler2} A. A. Burkov, M. D. Lukin, and E. Demler, Phys. Rev. Lett. \textbf{98} 200404 (2007). 

\bibitem{cardy} P. Calabrese and J. Cardy,  J. Stat. Mech. P06008 (2007).

\bibitem{sachdev} K. Sengupta, S. Powell and S. Sachdev, Phys. rev. A \textbf{69} 053616 (2004).

\bibitem{polkovnikov} V. Gritsev and A. Polkovnikov, eprint arxiv: 0910.3692.

\bibitem{chin} C.L.Hung, X.Zhang, N. Gemelke, and C.Chin eprint arxiv: 1003.0855.

\bibitem{demler} L.M.Duan, E.Demler and M.D.Lukin, Phys.Rev.Lett., \textbf{91}, 090402 (2003).

\bibitem{kuklov} A.B.Kuklov and B.V.Svistunov, Phys.Rev.Lett., \textbf{90}, 100401 (2003).

\bibitem{Leo} L. De Leo, C. Kollath, A. Georges, M.Ferrero, O. Parcollet, Phys. Rev. Lett., \textbf{101} 210403 (2008).

\bibitem{chaikin} P.M.Chaikin, and T.C.Lubensky, \textit{Principles of Condensed Matter Physics}, Cambridge University Press, (1995).

\bibitem{gemelke} C-L. Hung, X. Zhang, N. Gemelke and C. Chin, eprint arXiv: 1003.0855.

\bibitem{halperin} B. I. Halperin and P. Hohenberg, Phys. Rev. \textbf{188} 2, (1969).

\bibitem{wessel} M.Troyer, F. Alet, and S. Wessel, Braz. J. of Physics \textbf{34} 377 (2004).


\end{thebibliography}
\end{document}